\documentclass[a4paper,12pt]{article}
\usepackage{graphicx}
\usepackage{amsmath}
\usepackage{amssymb}
\usepackage{multirow}
\usepackage{float} 
\usepackage{cite}

\begin{document}

\begin{center}
{\large\bf  LHC signals of the next-to-lightest scalar Higgs state of the NMSSM in the $ 4\tau$ decay channel} \\
\vspace*{1.0truecm}
{\large M. M. Almarashi} \\
\vspace*{0.5truecm}
{\it Physics Department, Faculty of Science, Taibah University\\
P. O. Box 344, Medina, KSA}
\end{center}

\begin{abstract}
We study the $a_1a_1$ and $Za_1$ decay channels of the next-to-lightest CP-even Higgs boson $h_2$ of the NMSSM at the LHC, where the $h_2$
is produced in gluon fusion. It is found that while the $h_2$ discovery is impossible through the latter channel, the former one in the $ 4\tau$ final state
 is a promising channel to discover the $h_2$ with masses up to around 250 GeV at the LHC.
 Such a discovery of the $h_2$ is mostly accompanied with a light $a_1$, which is a clear evidence for 
distinguishing the NMSSM from the MSSM since such 
a light $a_1$ is impossible in the MSSM.
\end{abstract}

\section{Introduction}
\label{sect:intro}

The discovery of the standard model (SM)-like Higgs boson with a mass around 125 GeV at the LHC \cite{CMS1-Higgs,ATLAS1-Higgs,CMS2-Higgs,ATLAS2-Higgs} 
can be accommodated in the framework of the next-to-minimal supersymmetric standard model
(NMSSM) \cite{NMSSM1,NMSSM2,NMSSM3,NMSSM4,NMSSM5,NMSSM6,NMSSM7,NMSSM8,NMSSM9,upper,NMSSM10,NMSSM11} without much fine tuning, and as a consequence
it has acquired increasing attention. In this model one Higgs singlet field is added to the two MSSM-type Higgs
doublets in order to give a natural explanation of the $\mu$-problem of the MSSM \cite{Kim:1983dt}. So, the Higgs sector of the NMSSM is phenomenologically richer than that 
of the MSSM due to the existence of this extra Higgs singlet.

The Higgs spectrum of the NMSSM after electroweak symmetry breaking contains seven Higgs mass states,
assuming CP-conservation: two pseudoscalar Higgses $a_{1, 2}$ ($m_{a_1} < m_{a_2} $),
three scalar Higgses $h_{1, 2, 3}$ ($m_{h_1} < m_{h_2} < m_{h_3}$)
 and a pair of charged Higgses $h^{\pm}$. Following the discovery of the SM-like Higgs boson in 2012, the observation of
 additional Higgs bosons, if they exist, 
would point to the existence of supersymmetric extensions of the SM. In the NMSSM framework, the search for light Higgs bosons has been done by
many authors, aiming to establish the so-called a `no-lose theorem' of the NMSSM stating that one or more
of the Higgs bosons of the NMSSM should be discovered at the LHC throughout the entire NMSSM
parameter space \cite{NoLoseNMSSM1,NoLoseNMSSM2,NoLoseNMSSM3,NoLoseNMSSM4,NoLoseNMSSM5,NoLoseNMSSM6,NoLoseNMSSM7,
NoLoseNMSSM8,NoLoseNMSSM9,NoLoseNMSSM10}. All these studies were performed before the discovery of the Higgs boson at the LHC in 2012.
Many studies have also been done on the discovery potential of other Higgs bosons of the NMSSM following the 2012 discovery 
\cite{Kang:2013rj,Cerdeno:2013cz,Cao:2013si,Christensen:2013dra,Ellwanger:2013ova,Cerdeno:2013qta,King:2014xwa,Bomark:2014gya,
Chakraborty:2015xia,Ellwanger:2015uaz,Ellwanger:2016qax,Conte:2016zjp,Wang:2015omi,Guchait:2016pes,Das:2016eob,Beskidt:2017dil,
Almarashi:2018ign,Almarashi:2018rkn,Heng:2018kyd,Das:2017mqw,Das:2018fog,Baum:2019uzg,Wang:2020tap,Almarashi:2020fdp}.

One of the interesting feature of the NMSSM is that Higgs-to-Higgs decays are dominant over large regions of parameter space if they are kinematically allowed.
The importance of such decays in the framework of the NMSSM has long been emphasized in the literature, see, e.g.,
Ref.~\cite{Dermisek:2005ar}.
 It was found that Vector Boson Fusion (VBF)
 could be a suitable production channel to detect $h_{1,2}\to a_1a_1$ at the LHC,
 in which the Higgs pair decays into $jj\tau^+\tau^-$ \cite{NoLoseNMSSM2}. Both the Vector Boson Fusion and Higgs-strahlung production mechanisms could also be useful to discover
 such Higgses in the $4\tau$ final states \cite{Belyaev:2008gj}. Furthermore, some scope could be afforded by  $4\mu$ and $2\tau2\mu$ signatures 
 in the gluon-fusion production channel \cite{NoLoseNMSSM5,2mu2tau}. Higgs production in association with a $ b\bar b$ pair could also be a good means
 to search for the $h_{1,2}\to Za_1$ at the LHC \cite{Almarashi:2011qq}. All these studies were performed prior to the Higgs discovery in 2012.
 
  In this paper, we study the LHC discovery potential for   
 the next-to-lightest CP-even Higgs boson $h_2$, which is not a SM-like Higgs, decaying either into two 
light CP-odd Higgs bosons $a_1a_1$ or into a light $a_1$ and a gauge boson $a_1Z$
 through the gluon fusion $gg\to h_2$  in the $4\tau$ final state in the NMSSM framework.
 We calculate the
signal rates of the two processes $gg\to h_2\to a_1a_1 \to 4\tau$
and $gg\to h_2\to Za_1 \to 4\tau$  to examine whether or not there are some
regions of NMSSM parameter space where the $h_2$ and $a_1$ states can  simultaneously be observed at
the LHC.\footnote{We do not consider the case of both $ b\bar bb\bar b$ and $b\bar b\tau^+\tau^-$ final states because
these channels are burdened by large SM backgrounds.} We perform a partonic signal-to-background analysis of the $h_2$ production.
 It is found that there are parameter regions of the NMSSM where
the  $h_2$ and $a_1$ signals may be found at the LHC through the process $gg\to h_2\to a_1a_1 \to 4\tau$. 

The paper is planned as follows. In the next section we briefly discuss the Higgs sector of the NMSSM, describing the NMSSM parameter 
space scans performed under current constraints. In section ~\ref{sect:rates} we present 
the inclusive production rates of the $h_2$ 
at the LHC in the $4\tau$ final states as well as signal-to-background analysis for some benchmark points.
Finally, conclusions are given in section ~\ref{sect:summa}.

\section{The Higgs sector of the NMSSM}
The scale invariant superpotential of the NMSSM  in terms of the usual two MSSM-type Higgs
doublets superfields $\hat H_u$ and $\hat H_d$ as well as the singlet one $\hat S$ is given by \cite{NMSSM4,NMSSM5} 
 \begin{equation}  
 W_{NMSSM} = {\rm MSSM ~ Yukawa ~ terms} + \lambda\hat S\hat H_u\hat H_d 
+\frac{1}{3}\kappa{\hat S}^3,
\end{equation}
where both $\lambda$ and $\kappa$ are dimensionless couplings.
The term $\lambda\hat S\hat H_u\hat H_d$ is introduced to solve the $\mu$-problem of the MSSM superpotential. 
 When the singlet superfield develops a vacuum expectation value (VEV) $\langle S\rangle = \frac{1}{\sqrt{2}}\upsilon_s$
 upon spontaneous symmetry breaking, an `effective' $\mu$-parameter given by
 $\mu_{\rm eff} = \lambda\langle S\rangle$ of the order of the electroweak scale will be automatically generated.
The last term of the above equation is introduced to
avoid the Peccei-Quinn symmetry \cite{Peccei:1977hh, Peccei:1977ur}.
The soft breaking terms for both the doublet and singlet fields read
\begin{equation}
V_{\rm NMSSM}=m_{H_u}^2|H_u|^2+m_{H_d}^2|H_d|^2+m_{S}^2|S|^2
             +\left(\lambda A_\lambda S H_u H_d + \frac{1}{3}\kappa A_\kappa S^3 + {\rm h.c.}\right),
\end{equation}
where $A_\lambda$ and $A_\kappa$ are the trilinear coupling parameters of the order of SUSY mass scale $m_{\rm{SUSY}}$.

The physical Higgs bosons arise after the Higgs fields acquire vacuum expectation values (VEVs), $\langle H_u\rangle =\frac{1}{\sqrt{2}} \upsilon_u$, 
 $\langle H_d\rangle = \frac{1}{\sqrt{2}} \upsilon_d$ and $\langle S\rangle = \frac{1}{\sqrt{2}}\upsilon_s$,
and eliminating the Goldstone boson states . As a result,
the potential has terms for the non-zero mass modes for the scalar fields $S_i (i = 1, 2, 3)$, pseudoscalar fields $P_i (i = 1, 2)$ and charged fields $h^\pm$ given by
\begin{equation}
V_{mass}= \frac{1}{2}(S_1\ \   S_2\ \ S_3)\mathcal{M}_S\left( \begin{array}{ccc}S_1 \\ S_2 \\ S_3 \end{array} \right)
 +\frac{1}{2}(P_1\ \   P_2)\mathcal{M}_P\left( \begin{array}{ccc}P_1 \\ P_2 \end{array} \right)
 + m^2_{h^\pm}h^+h^-.
\end{equation}
One can obtain physical mass eigenstates with tree-level masses as follows.  First, the elements of the mass matrix
for the CP-even Higgs states at tree-level are given by \cite{MNZ}
\begin{equation}
 \mathcal{M}_{S11}=m^2_A+\bigg (m^2_Z-\frac{1}{2}(\lambda\upsilon)^2\bigg ){\rm sin}^22\beta,
\end{equation}
\begin{equation}
 \mathcal{M}_{S12}=-\frac{1}{2}\bigg (m^2_Z-\frac{1}{2}(\lambda\upsilon)^2\bigg ){\rm sin}4\beta,
\end{equation}
\begin{equation}
\mathcal{M}_{S13}=-\frac{1}{2}\bigg (m^2_A{\rm sin}2\beta+2\frac{\kappa{\mu^2_{\rm eff}}}{\lambda}\bigg )\bigg (\frac{\lambda\upsilon}{\sqrt{2}\mu_{\rm eff}}\bigg ){\rm cos}2\beta 
\end{equation}
\begin{equation}
 \mathcal{M}_{S22}=m^2_Z{\rm cos}^22\beta+\frac{1}{2}(\lambda\upsilon)^2{\rm sin}^22\beta,
\end{equation}
\begin{equation}
 \mathcal{M}_{S23}=\frac{1}{2}\bigg (4{\mu^2_{\rm eff}}-m^2_A{\rm sin}^22\beta-\frac{2\kappa{\mu^2_{\rm eff}}{\rm sin}2\beta}{\lambda}\bigg )\frac{\lambda\upsilon}{\sqrt{2}\mu_{\rm eff}}
\end{equation}
\begin{equation}
\mathcal{M}_{S33}=\frac{1}{8}m^2_A{\rm sin}^22\beta\frac{\lambda^2\upsilon^2}{\mu^2_{\rm eff}}+4\frac{\kappa^2{\mu^2_{\rm eff}}}{\lambda^2}
+\frac{\kappa A_\kappa \mu_{\rm eff}}{\lambda}-\frac{1}{4}\lambda\kappa\upsilon^2{\rm sin}2\beta,
\end{equation}
where 
 $m^2_A=\sqrt{2}\frac{\mu_{\rm eff}}{\sin2\beta}\bigg(A_{\lambda}+\frac{\kappa\mu_{\rm eff}}{\lambda}\bigg)$,
tan$\beta=\frac{\upsilon_u}{\upsilon_d}$ and $\upsilon^2 = {\upsilon^2_u} + {\upsilon^2_d} $.

Second, the elements of the mass matrix for the CP-odd Higgs states at tree-level are \cite{MNZ}
\begin{equation}
\mathcal{M}_{P11}=m^2_A,
\end{equation}
\begin{equation}
 \mathcal{M}_{P12}=\frac{1}{2}\bigg (m^2_A{\rm sin}2\beta-6\frac{\kappa{\mu^2_{\rm eff}}}{\lambda}\bigg )\frac{\lambda\upsilon}{\sqrt{2}\mu_{\rm eff}},
\end{equation}
\begin{equation}
\mathcal{M}_{P22}=\frac{1}{8}\bigg (m^2_A{\rm sin}2\beta+6\frac{\kappa{\mu^2_{\rm eff}}}{\lambda}\bigg )\frac{\lambda^2\upsilon^2}{\mu^2_{\rm eff}}{\rm sin}2\beta
-3\frac{\kappa\mu_{\rm eff}A_\kappa}{\lambda}.
\end{equation}

Third, the mass of charged Higgs fields at tree-level is given by \cite{MNZ}
\begin{equation}
 m^2_{h^\pm}=m^2_A+m^2_W-\frac{1}{2}(\lambda \upsilon)^2.
\end{equation}

It is clear from the above equations that the Higgs sector of the NMSSM at tree-level is described by the six parameters: $\lambda$, $\kappa$, tan$\beta$, $\mu_{\rm eff}$ , $A_\lambda$ and 
$A_\kappa$. Assuming CP-conservation, the upper mass bound for the lightest CP-even Higgs boson of the NMSSM, if it is the SM-like Higgs,
at tree level is given by \cite{NMSSM4,NMSSM5}
\begin{equation}
 m^2_{h_1} < m^2_Z {\rm cos}^2(2\beta)+\frac{\lambda^2 \upsilon^2}{2}{\rm sin}^2(2\beta).
\end{equation}
The last term in this expression can lift $m_{h_1}$ with up to 10-15 GeV higher than the corresponding one of the MSSM. So,
less loop corrections are required to get the lightest Higgs, $h_1$, to be SM-like with mass around 125 GeV.
Clearly, large values of $\lambda$ and low values of tan$\beta$ are preferred to obtain a large value of the $h_1$ at tree level.
The scenarios with $m_{h_1} < 125$ GeV means that the $h_1$ is highly singlet-like so it can escape the constraints coming from LEP, the Tevatron
and the LHC. In this case, the next-to-lightest CP-even Higgs boson $h_2$ is the SM-like Higgs of mass around 125 GeV.

In this paper we are interested in the production of the next-to-lightest scalar Higgs boson $h_2$, which is not the SM-like Higgs, and its decays into either two 
light CP-odd Higgs bosons, $a_1a_1$, or a light CP-odd Higgs and a gauge boson, $a_1Z$, in the mass region $m_{h_2} \lesssim 300$ GeV. We use
the package NMSSMTools5.1.2 \cite{NMHDECAY1,NMHDECAY2,NMSSMTools} which computes the masses, couplings and decay widths of all the 
Higgs bosons in addition to the spectrum of supersymmetric particles. This package takes into account various theoretical and experimental constraints such as
constraints from negative Higgs searches at LEP, the Tevatron and the LHC,
as well as SUSY mass limits as implemented in the package.
Moreover, it takes into account constraints of Upsilon, B and K decays and also the bounds on the mass of the SM-like Higgs and its signal
rates. More details about the constraints can be found on the website of the package.
We have ignored the constraints on the dark matter because its nature is still unknown. We also do not take into account 
the constraints on the muon anomalous magnetic moment because such constraints have large theoretical uncertainties.

In our parameter space, we scan by varying the tree level parameters of the NMSSM within the following ranges:
\begin{center}
$0.6 \leq \lambda \leq  0.7$, \phantom{aa} $-0.65 \leq \kappa \leq  0.65$,\phantom{aa} $1.6 \leq \tan\beta \leq  60$, \phantom{aa} \\
$100 \leq \mu_{\rm eff} \leq  200$ GeV, \phantom{aa} $-2000 \leq A_{\lambda} \leq  2000$ GeV,\phantom{aa} $-10 \leq A_{\kappa} \leq  10$ GeV. \\
\end{center}
Notice that we focus here on scenarios with large values of $\lambda$ and small values of both $\mu_{\rm eff}$ and $A_{\kappa}$ to simultaneously obtain
the $h_2$ with $m_{h_2} \lesssim 300$ GeV and a light $a_1$. Remaining soft mass parameters for the scalars and gauginos in addition to
the trilinear soft SUSY coupling parameters, contributing at higher order level, are set to \\
$\bullet\phantom{a}m_Q = m_U = m_D = m_L = m_E = m_{Q_3} = m_{U_3} = m_{D_3} = m_{L_3} = m_{E_3} = 3000$ GeV,\\
$\bullet\phantom{a} M_1 = 500$ GeV, $M_2 = 1000$ GeV,  $M_3 = 3000$ GeV,\\
$\bullet\phantom{a}A_{U_3} = A_{D_3} = A_{E_3} = 3000$ GeV.\\

We randomly perform a scan over the above mentioned parameters and identify the parameter space of the NMSSM that passed
all theoretical and experimental constraints. The outcome of our scan contains masses and branching ratios 
of the NMSSM Higgs bosons for all the surviving data points which have passed 
the constraints. As mentioned above, we have ignored the constraints on dark matter relic density. To know
the effects of these constraints on the NMSSM parameter space see, for example, Ref.\cite{Aldufeery:2020dkb} and
references therein.

\section{\large Higgs boson signal rates}
\label{sect:rates}

In this section we discuss the discovery potential of the $h_2$ produced in the gluon fusion $gg\to h_2$ at the LHC in the mass region $m_{h_2} \lesssim 300$ GeV.
The region with $m_{h_2} \gtrsim 300$ GeV is less promising since the cross sections for the $h_2$ production fall quickly with increasing $h_2$ masses.
For the surviving data points obtained from the random scan,
we calculate the inclusive cross sections for the $h_2$ production by using 
CalcHEP \cite{CalcHEP} for the following processes:
\begin{equation}
gg\to h_2\to a_1a_1\to \tau^+\tau^-\tau^+\tau^- \phantom{aa} {\rm and} \phantom{aa}   gg\to h_2\to Za_1\to \tau^+\tau^-\tau^+\tau^-.
\label{eq:proc}
\end{equation}
Here, we consider a center-of-mass 
energy $\sqrt s=14$ TeV for the LHC. 

\begin{figure}
 \centering\begin{tabular}{cc}
 \includegraphics[scale=0.6]{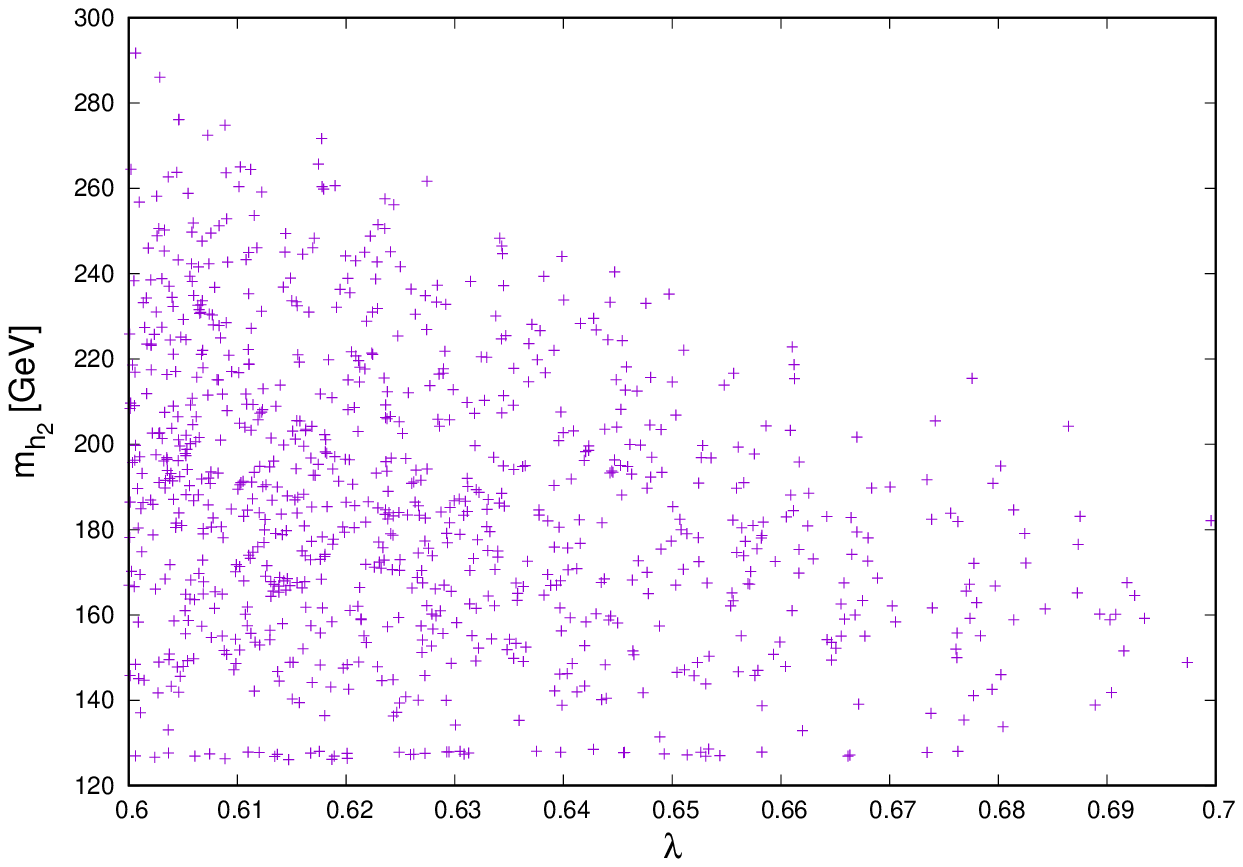} &\includegraphics[scale=0.6]{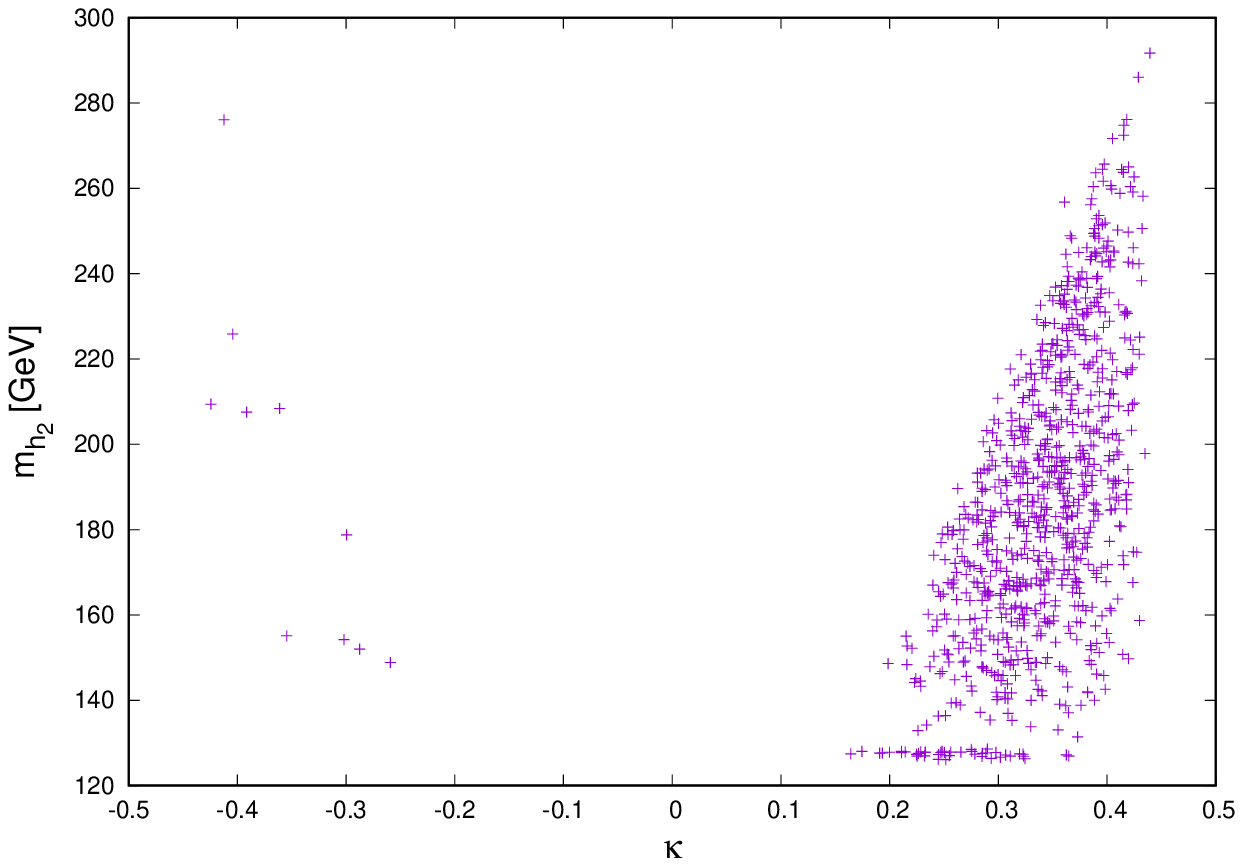}\\
 \includegraphics[scale=0.6]{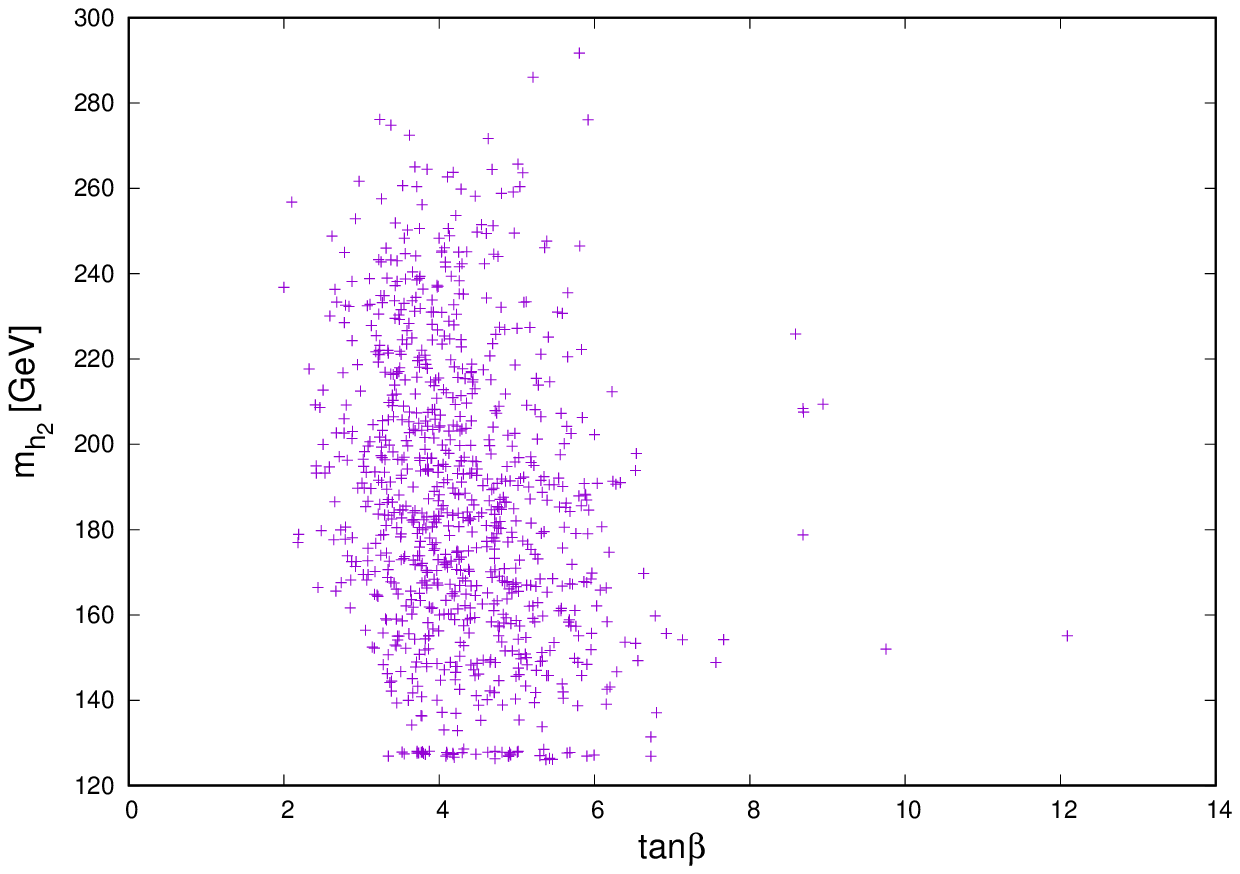} &\includegraphics[scale=0.6]{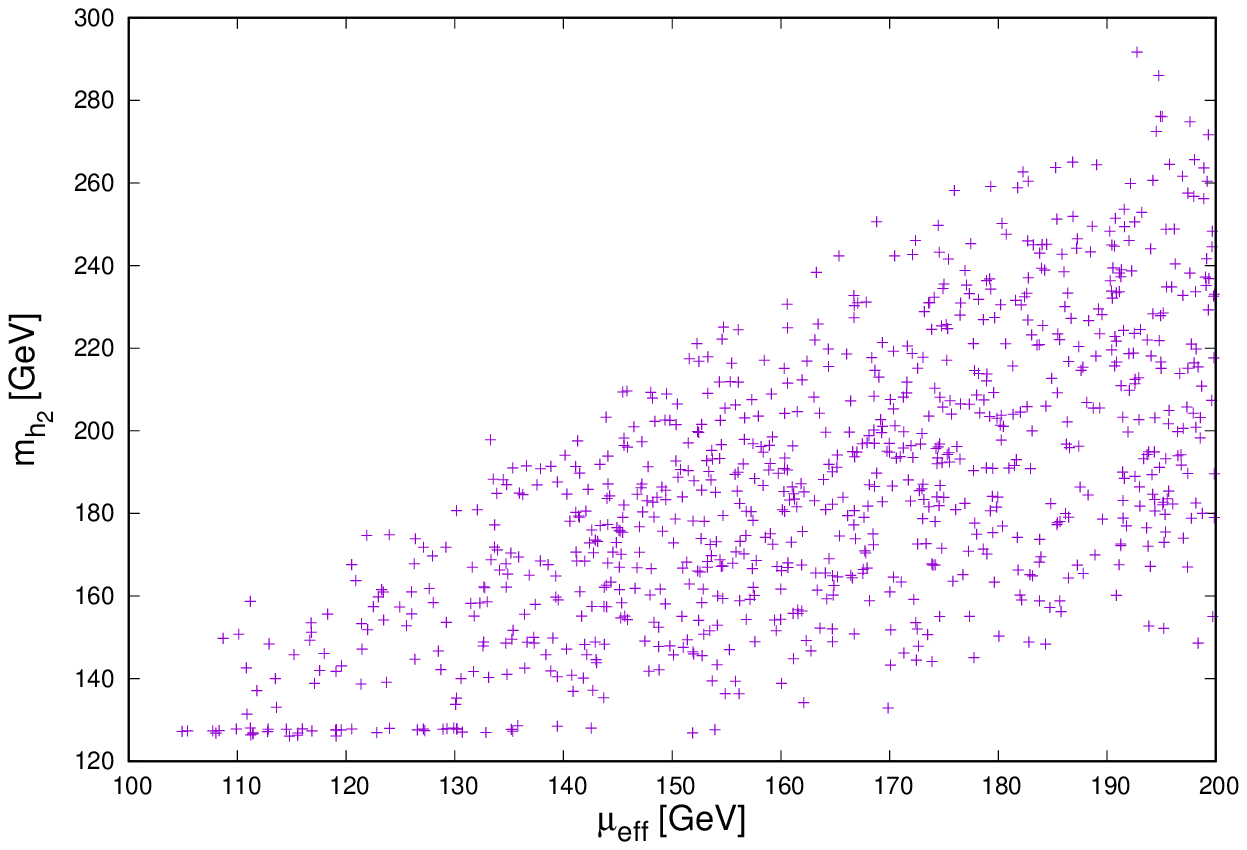}\\
 \includegraphics[scale=0.6]{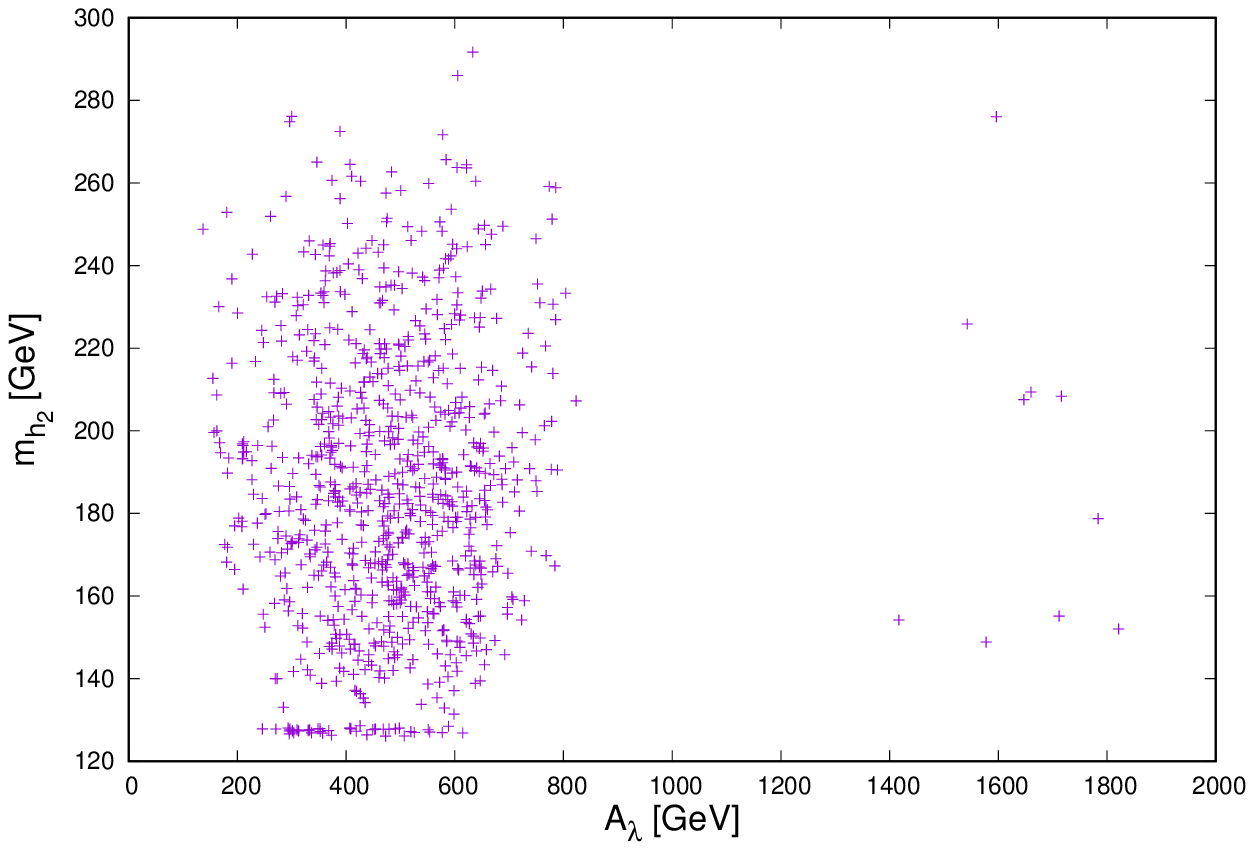} & \includegraphics[scale=0.6]{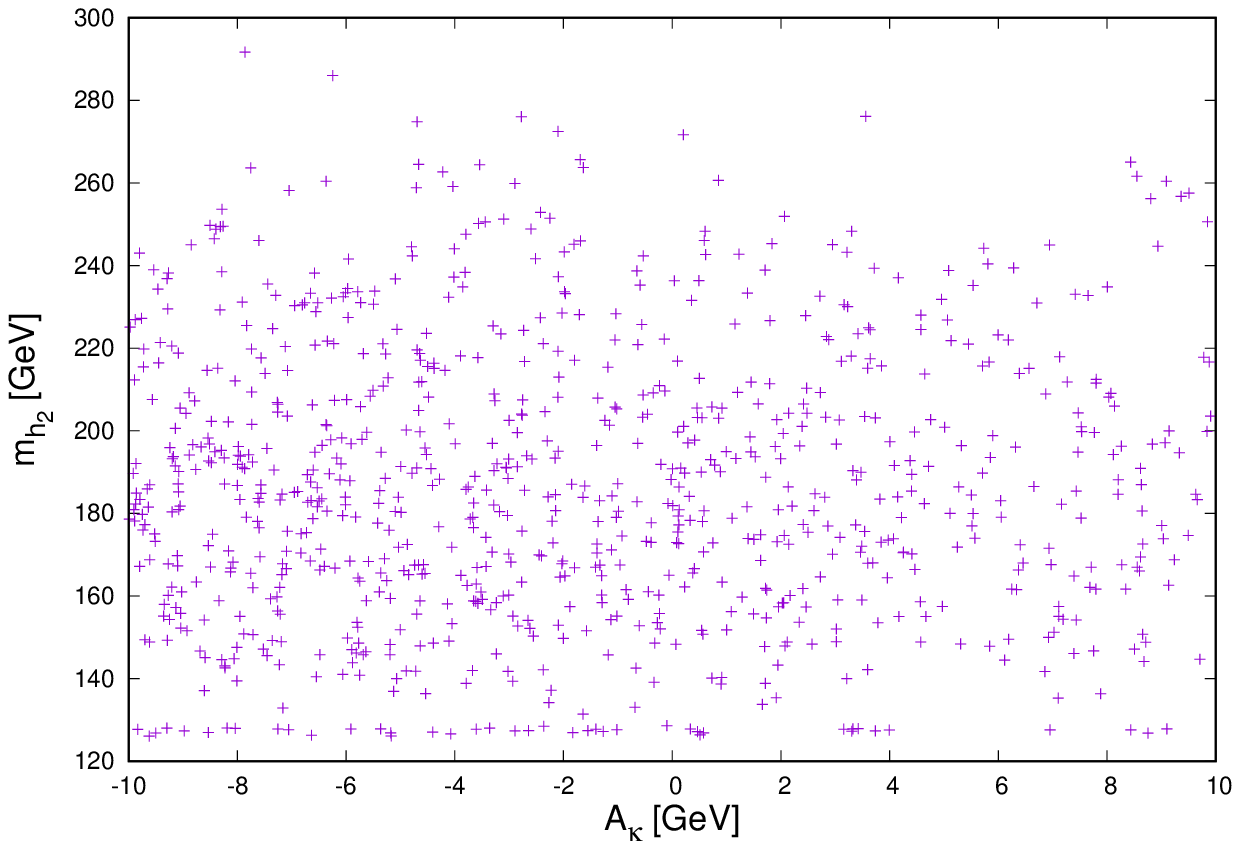}
 \end{tabular}

\caption{The next-to-lightest CP-even Higgs mass $m_{h_2}$ as a function of $\lambda$ and $\kappa$ (top-panels), $\tan\beta$ and $\mu_{\rm eff}$ (middle-panels)
and  $A_\lambda$ and $A_\kappa$ (bottom-panels).}
  \label{fig1}
\end{figure}

In Fig.~\ref{fig1} we present the mass of the next-to-lightest CP-even Higgs boson  $m_{h_2}$  against the tree-level parameters of the NMSSM.
One finds that for the surviving points the $m_{h_2}$ decreases by increasing $\lambda$ whereas it increases by increasing both $\kappa$ and $\mu_{\rm eff}$.
Also, it is found that for our parameter space most of the surviving points correspond to the region with $2\lesssim \tan\beta \lesssim 7$ and 
$200 \lesssim A_{\lambda}  \lesssim 800$ while the distribution in $A_\kappa$ is quite uniform.

\begin{figure}
 \centering\begin{tabular}{c}
 \includegraphics[scale=0.8]{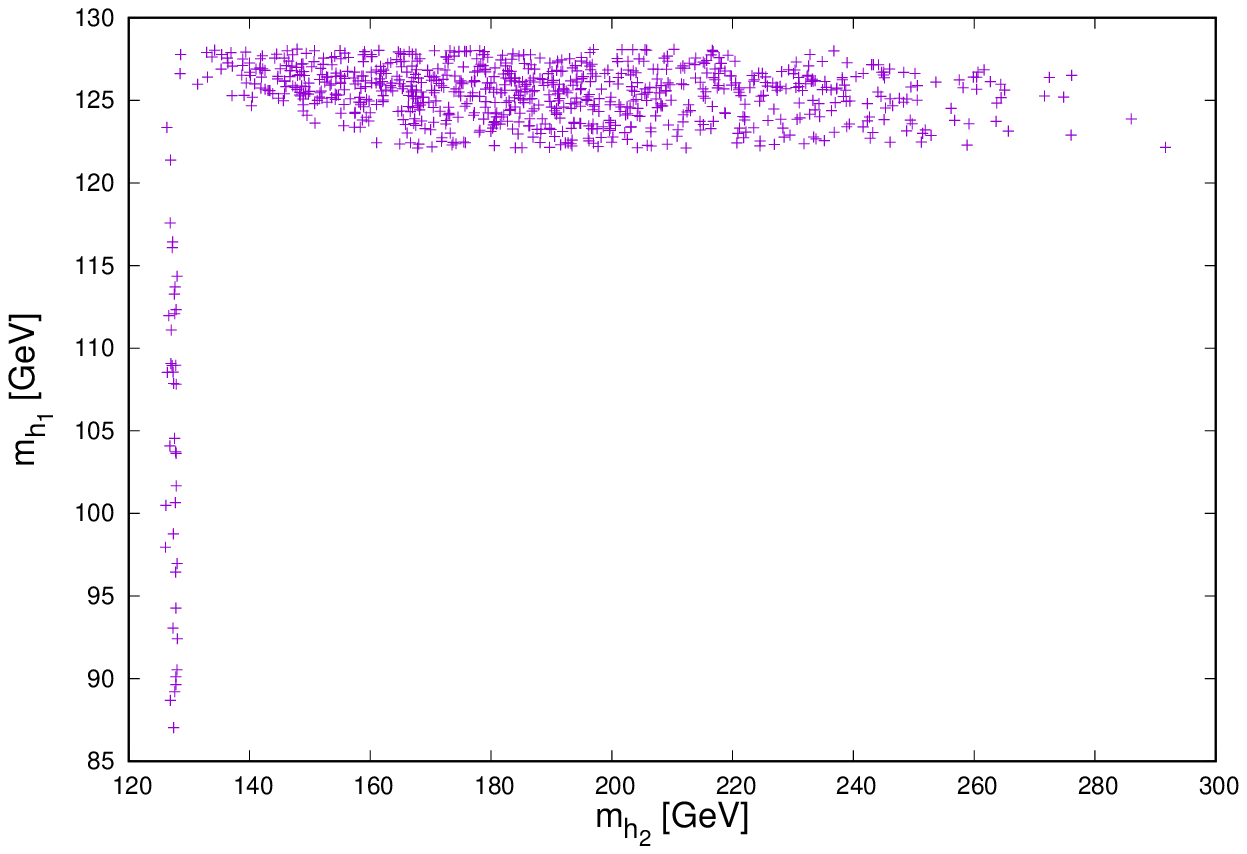}

 \end{tabular}

\caption{The mass distribution for the next-to-lightest CP-even Higgs, $m_{h_2}$,
versus the lightest CP-even Higgs mass, $m_{h_1}$. }
\label{fig2a}
\end{figure}

\begin{figure}
 \centering\begin{tabular}{c}
 \includegraphics[scale=0.8]{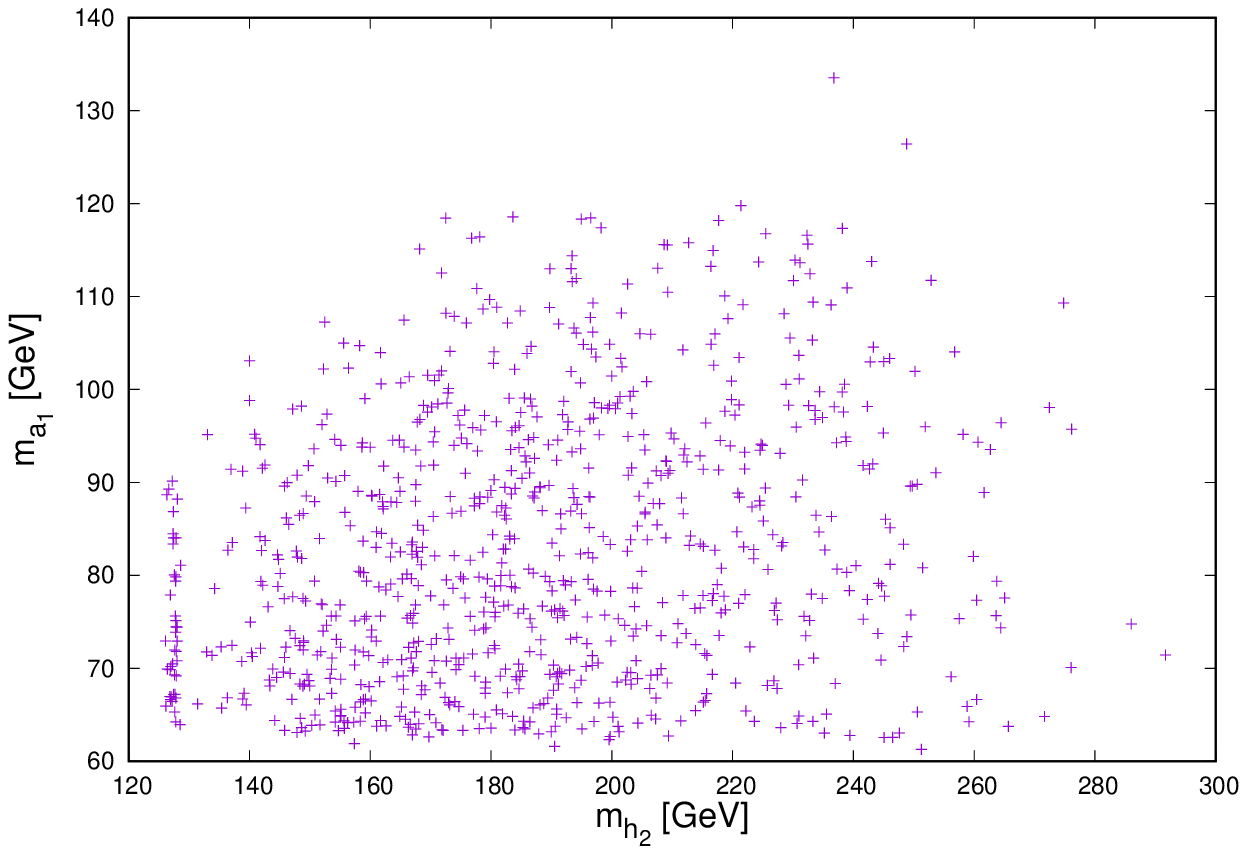}

 \end{tabular}

\caption{The mass distribution for the next-to-lightest CP-even Higgs, $m_{h_2}$,
versus the lightest CP-odd Higgs mass, $m_{a_1}$. }
\label{fig2b}
\end{figure}

Looking at Figs.~\ref{fig2a} and ~\ref{fig2b}, showing the correlations between $m_{h_2}$ and 
the lightest CP-even Higgs mass $m_{h_1}$ and between $m_{h_2}$ and the lightest CP-odd Higgs mass $m_{a_1}$ for the surviving points,
it is clear that in most regions of our parameter space the $h_1$ is the SM-like Higgs and the $h_2$ can only play the role of the SM-like
Higgs in a small region of the parameter space. The $h_2$ is a mixture of doublet and singlet components for
the majority of points selected in our parameter space. Fig.~\ref{fig2b}  shows 
that the smaller $m_{a_1}$ the smaller $m_{h_2}$. Since the surviving points  have small
values of $A_{\kappa}$, only small values of $m_{a_1}$ are allowed. One noteworthy feature of the figure is that the $h_2$ can be the SM-like Higgs with mass 
around 125 GeV, which corresponds to a light $a_1$ with $m_{a_1} \lesssim 90$ GeV.

\begin{figure}
 \centering\begin{tabular}{cc}
 \includegraphics[scale=0.6]{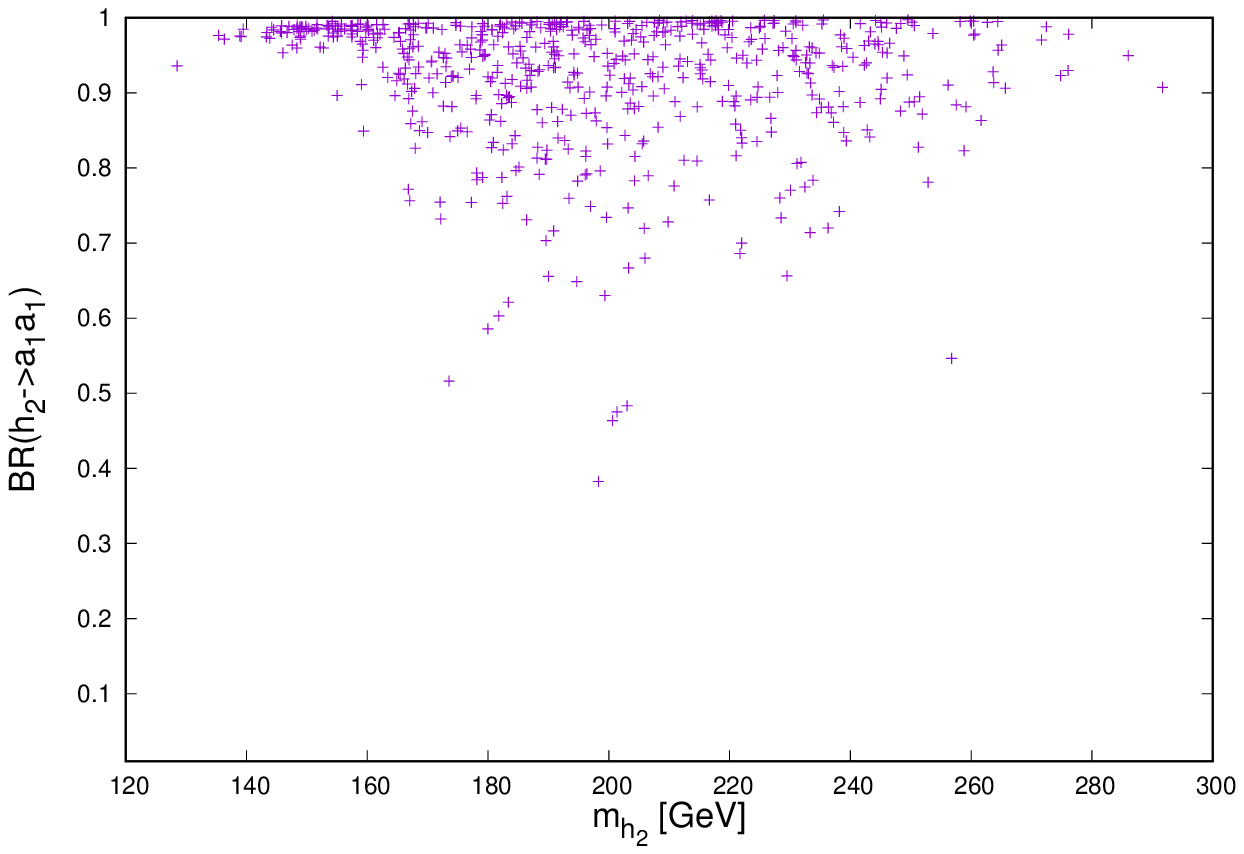} &\includegraphics[scale=0.6]{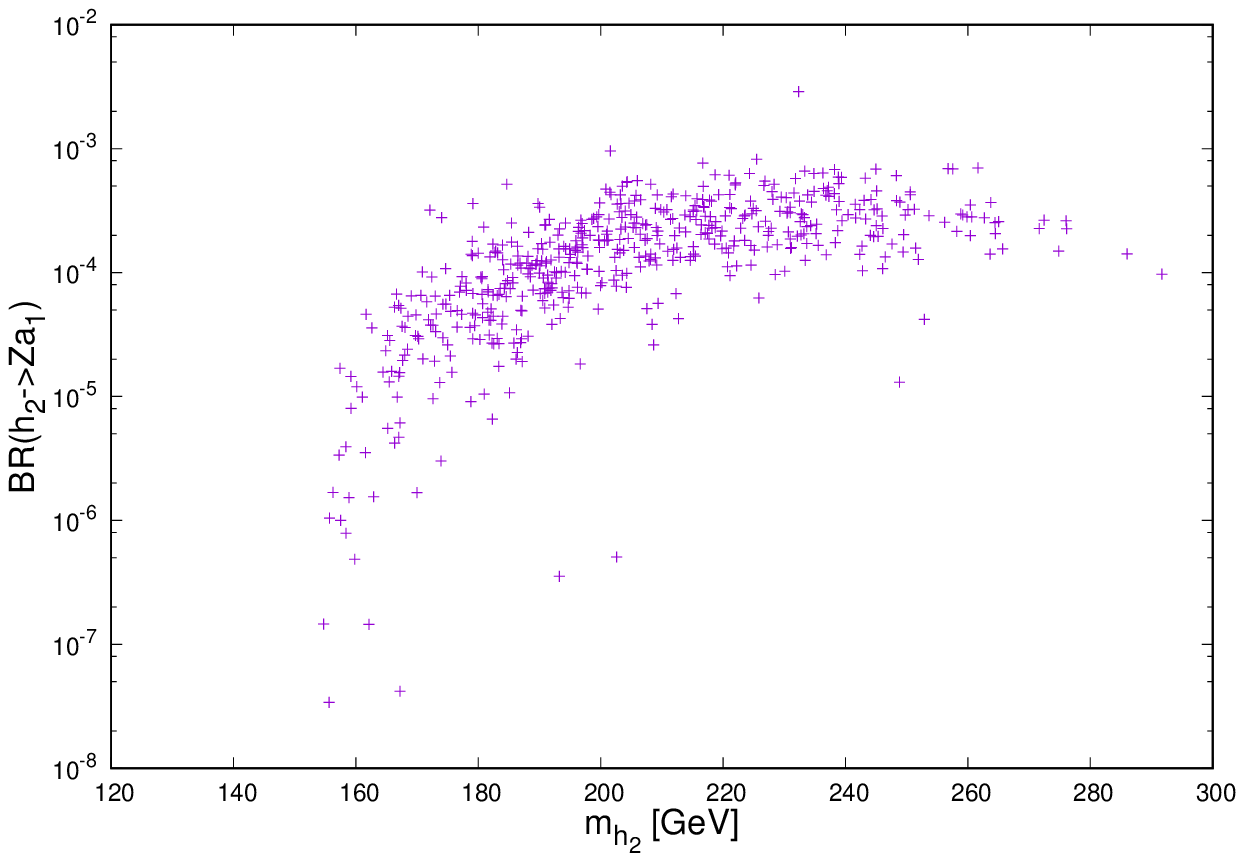}
 \end{tabular}

\caption{The next-to-lightest CP-even Higgs mass $m_{h_2}$ plotted against both
 $~{\rm BR}(h_2\to a_1a_1)$ (left) and $~{\rm BR}(h_2\to Za_1)$ (right).}
\label{fig3}
\end{figure}

\begin{figure}
 \centering\begin{tabular}{cc}
 \includegraphics[scale=0.6]{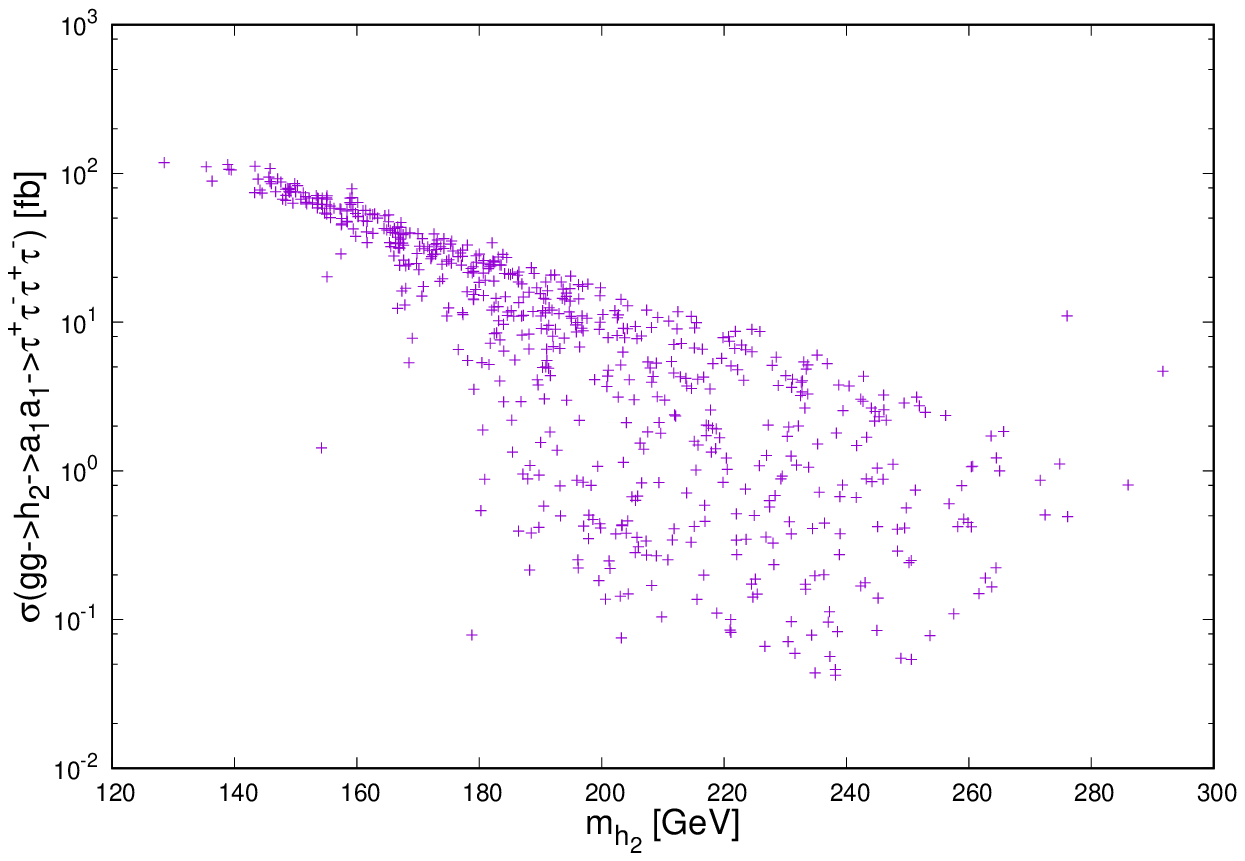} &\includegraphics[scale=0.6]{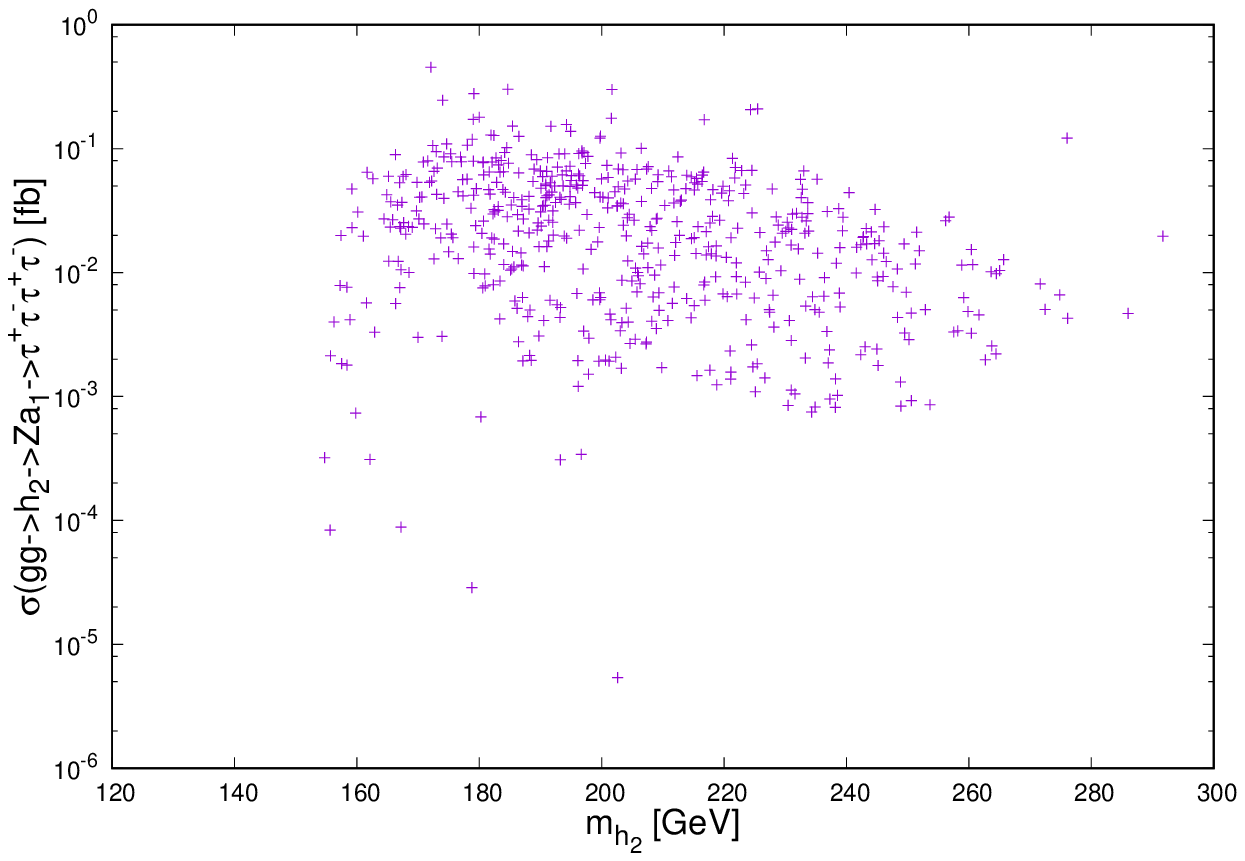}
 \end{tabular}

\caption{The production rates in fb for $\sigma(gg\to h_2).{\rm BR}(h_2\to a_1a_1\to \tau^+\tau^-\tau^+\tau^-$) (left) and 
 $\sigma(gg\to h_2).{\rm BR}(h_2\to Za_1\to \tau^+\tau^-\tau^+\tau^-$) (right)
as functions of $m_{h_2}$.}
\label{fig4}
\end{figure}

Due to the mixing between the Higgs doublets and singlet, Higgs-to-Higgs decays are
kinematically allowed for large area of the NMSSM parameter space, even for small
masses of Higgs bosons. Also, one distinguished landmark of the NMSSM is that the existence
of the lightest CP-odd Higgs boson $a_1$ with mass values less than $m_Z$ is quite natural,
 which is impossible in the context of the MSSM.
In Fig.~\ref{fig3} we display the correlations between $m_{h_2}$ and the $h_2$ decays into light CP-odd Higgs pairs $h_2\to a_1a_1$ (left)
and into an $a_1$ and a gauge boson $Z$ (right). It is clear from the left-panel of the figure that the decay $h_2\to a_1a_1$ is
the dominant one whenever it is kinematically open. It is found that the BR($h_2\to a_1a_1$) ranges from around 40\% to 100\%, see the left-panel of the figure, 
while the maximum BR($h_2\to Za_1$) is 0.1\%, see the right-panel of the figure. In fact, the dominance of $a_1a_1$ decay channel
causes a suppression to other decay channels such as $b\bar b$ and other channels.

Fig.~\ref{fig4} illustrates the inclusive $h_2$ production rates ending up with $a_1a_1\to \tau^+\tau^-\tau^+\tau^-$ (left) and
$Za_1\to \tau^+\tau^-\tau^+\tau^-$ (right) as functions of $m_{h_2}$.
It is shown from the figure that the $h_2$ production rates decrease rapidly with increasing $m_{h_2}$. It is clear that the production rates 
$\sigma(gg\to {h_2} \to a_1a_1\to \tau^+\tau^-\tau^+\tau^-)$ are sizable, reaching up to 100 fb for small values of $m_{h_2}$ while
the production rates $\sigma(gg\to {h_2} \to Za_1\to \tau^+\tau^-\tau^+\tau^-)$ is quite small, reaching around 0.5 fb at the most. The latter production rates
are clearly not enough to detect the $h_2$ at the LHC, taking into account that leptonic tau decays are around 17.5\%.
In short,
the inclusive production rates for $h_{ 2}$ decaying into $a_1a_1\to \tau^+\tau^-\tau^+\tau^- $ are promising and could allow discovery of both $h_2$ and $a_1$ at the LHC
while the production rates for $h_2$ decaying into  $Za_1\to \tau^+\tau^-\tau^+\tau^-$ are quite small. So, we will analyze signal-to-background for the former channel
as the  latter channel is useless.

To claim discovery at the LHC, we have done a partonic signal-to-background (S/B)
analysis based on CalcHEP. The dominant standard model backgrounds are  the irreducible background coming from
$pp\to  \tau^+\tau^- \tau^+\tau^-$(via $\gamma$ and $Z$ exchange). Here, we assume 
 the double leptonic decay channels of the $\tau$'s. In the table 1, we give 4 points as benchmark points for various masses of the $h_2$ to do our analysis.
 Here, we assume that the tagging efficiency is 50$\%$ for tau -jets, by scaling of the total
cross-sections.
As it is shown in the table we have calculated the cross sections for both the signal and background processes and also the significance $S/\sqrt B$ at
a center-of-mass energy $\sqrt s=14$ TeV for the LHC. We have done that for 300 fb$^{-1}$ and 1000 fb$^{-1}$ of accumulated
luminosity.\footnote{We do such analysis without assuming any cuts. Of course, the proper cuts could improve the signal to background ratio, which we leave 
for the experimentalist to do.}
It is obvious that there is a good potential
to detect the $h_2$ decaying into $a_1a_1\to \tau^+\tau^-\tau^+\tau^- $ at the LHC in the mass region $140 \lesssim m_{h_2} \lesssim 220$ GeV.
The corresponding signal events are quite large of order 31680 events for $m_{h_2} = 140$ GeV and 2364 events for  $m_{h_2} = 220$ events
with 300 fb$^{-1}$ of integrated luminosity. Again these results given without assuming any cuts, which of course reducing the number of signal rates.
Thus, we conclude that the LHC with integrated luminosity of 1000 fb$^{-1}$ has the potential to discover the $h_2$, if it is not a SM-like Higgs, with masses up to around 250 GeV.
Such a discovery of the $h_2$ is mostly accompanied with a light $a_1$. The existence of such a light $a_1$ is a direct evidence for 
distinguishing the NMSSM from the MSSM.

\begin{table}[h]
\caption{Four benchmark points P1, P2, P3, and P4 used in the $S/\sqrt B$ analysis.}
{\begin{tabular}{|c|c|c|c|c|}
\hline
 & P1 & P2 & P3 & P4  \\ 
\hline 
$\lambda$ & 0.615706 & 0.650828 &  0.637590 & 0.617789  \\
\hline
$\kappa$ &  0.261287 & 0.264725 &  0.339134 &  0.387478 \\
\hline
$\tan\beta$ & 5.2247 & 3.78738 & 3.82514 & 3.70979 \\
\hline
$\mu_{\rm eff}  {\rm [GeV]}$ & 153.678 & 198.766 & 198.201 & 199.224 \\
\hline
$A_\lambda {\rm [GeV]}$  &  646.778 & 517.464 & 464.215 & 426.835  \\
\hline
$A_\kappa {\rm [GeV]}$  & -8.00937 & 5.1126 & -9.72344 & 9.09329 \\
\hline
${\rm m_{h_2} [GeV]}$ & 140 & 180 & 220 & 260\\
\hline
${\rm m_{a_1} [GeV]}$ & 66 & 64 & 99 & 67 \\
\hline
$S$ [fb] with 300 fb$^{-1}$ & $3.168 \times 10^4$ & $8.61 \times 10^3$ & $2.364 \times 10^3$ & $3.21 \times 10^2$ \\
\hline
$B $ [fb] with 300 fb$^{-1}$ & $3.9 \times 10^4$ & $3.9 \times 10^4$ & $3.9 \times 10^4$ & $3.9 \times 10^4$ \\
\hline
$S/\sqrt B$ with 300 fb$^{-1}$  & 160.4 & 43.6 & 12 & 1.6 \\
\hline
$S/\sqrt B$ with 1000 fb$^{-1}$  & 292.9 & 79.6 & 21.9 & 3 \\ 
\hline
\end{tabular}}

\label{t1}

\end{table}

\newpage

\section{Conclusions}
\label{sect:summa}
In this paper, we have explored the discovery prospects of the next-to-lightest CP-even Higgs state, $h_2$, at the LHC with
 $\sqrt{s} = 14$ TeV. We have studied the detectability of the $h_2$ in the two processes
 $gg\to h_2\to a_1a_1\to \tau^+\tau^-\tau^+\tau^- $ and $gg\to h_2\to Za_1\to \tau^+\tau^-\tau^+\tau^- $. We have shown that
 while the $h_2$ discovery of the latter channel is impossible due to smallness of the inclusive production rates, the former
 channel is promising as the $\sigma (gg\to h_2\to a_1a_1\to \tau^+\tau^-\tau^+\tau^-)$ is sizable,
 and should help discovering the $h_2$ signals with masses up to around 250 GeV at the LHC with integrated luminosity of 1000 fb$^{-1}$.
 
 After doing some analysis for signals and dominant backgrounds in the partonic level, we have proven that the discovery of 
 both the $h_2$ and $a_1$ are possible at the LHC. Such a discovery of the $h_2$ is mostly accompanied with a light $a_1$
 with $m_{a_1}\lesssim m_Z$. 
 The existence of such a light $a_1$ is a direct evidence for 
 the NMSSM as such a light $a_1$ is impossible in the MSSM. Of course, more experimental analysis including
 $\tau$-decays, detector effects, parton shower, and hadronization are needed  to claim the 
actual discovery potential of such Higgses at the LHC. However, we believe that our results are valuable for scientists
interested in determining the NMSSM Higgs signals at the LHC.

The discovery of Higgs states through the process $gg\to h_2\to a_1a_1\to \tau^+\tau^-\tau^+\tau^-$
 has in fact two  merits. On the one hand, it can be a good
 alternative to discover both $h_2$ and $a_1$ that could
be difficult to be discovered in direct production. On the other
hand, it can be exploited to measure the trilinear Higgs self-coupling $h_2a_1a_1$.

\section*{Acknowledgments}
We gratefully acknowledge support of Taibah University, KSA.

\end{document}